\documentclass[12pt]{iopart}
\usepackage{amssymb}
\usepackage{setstack}
\usepackage{graphicx}
\usepackage{amsbsy}

\begin{document}
\title[One dimensional Potts model with many-body interactions]{One dimensional Potts model with many-body interactions and the Generalized Model of Polypeptide Chain for the helix-coil transition}

\author{Artem Badasyan}
\address{Department of Theoretical Physics, J. Stefan Institute,\\ Jamova 39, SI-1000 Ljubljana, Slovenia, EU}
\ead{abadasyan@gmail.com}

\author{Achille Giacometti}
\address{Dipartimento di Scienze Molecolari e Nanosistemi, Universit\`a Ca' Foscari Venezia,
Calle Larga S. Marta DD2137, I-30123 Venezia, Italy, EU}

\author{Rudolf Podgornik}
\address{Department of Theoretical Physics, J. Stefan Institute and Department of Physics, Faculty of Mathematics and Physics, University of Ljubljana - SI-1000 Ljubljana, Slovenia, EU}

\author{Yevgeni Mamasakhlisov and Vladimir Morozov}
\address{Department of Molecular Physics, Yerevan State University,\\ A.Manougian Str.1, 375025, Yerevan, Armenia}

\date{\today}

\begin{abstract}
Helix-coil transition in polypeptides is an example of a spin model with a preferred spin direction, in the sense that a theoretical formulation of this problem requires to assign a preferred value of spin to the helical conformation in order to account for different symmetries of the helical {\sl vs.} the coil states. This leads to the spin Hamiltonian of the {\sl Generalized Model of Polypeptide Chain} (GMPC) variety as opposed to the Potts model variety, both with many-body interactions. We compare the explicit solution of the Potts model and the solution of the GMPC within the transfer-matrix formalism. Comparison of both secular equations reveals that the largest eigenvalue of the Potts model with $\Delta$ many-body interactions is identical to the largest eigenvalue of the GMPC model with $\Delta-1$ many-body interactions, indicating the equivalence of both free energies. In distinction, the second largest eigenvalues do not coincide, leading to different thermal behavior of the spatial correlation length, related to the helix-coil transition interval. Spin models with built-in spin anisotropy thus engender different physical properties in the thermodynamic limit that we explore in detail.
\end{abstract}


\maketitle
\section{Introduction}
\label{sec:introduction}

Effective one-dimensional spin models have been widely applied to the description of thermodynamic properties of hard condensed matter \cite{baxter,wu}. However, these models are in general not very useful since they do not result in a phase transition, known to exist in such systems. Nevertheless, the description of systems like quantum dots, quantum wires and then all the way to different soft matter systems, now opens up a possibility to revitalize the importance of one-dimensional spin models for the description of systems with low dimensionality, where one can envision the ordered state as being realized only when spins have a preferred direction. The application of Potts-like many valued spin models allows to distinguish this preferred direction by assigning it a selected spin value. Assuming only nearest neighbor interactions, one can construct a Hamiltonian preferring spin states where nearest neighbor spins are in the same, preferred orientation. This model would be different from the classical Potts model and it is not clear, to what extent the introduction of preferred spin value influences the physical properties. There are several soft matter systems where this kind of considerations would make sense, {\sl e.g.} a polypeptide undergoing helix-coil transition \cite{cantor, polsher} and the stretch-induced transformation from the standard B-DNA conformation to a more extended S-DNA form \cite{Bruinsma, Punkkinen}. 

Specifically, for the helix-coil transition the description of polypeptide conformations can be reduced to consideration of a pair of torsional angles, related to each of the peptide units \cite{flory}. A two-dimensional plot (Ramachandran's plot) of accessible {\sl vs.} not accessible regions of these variables shows, that helix formation is promoted only when both torsional angles assume values from a well-defined $\alpha$-helical region of values. When modeled in terms of spins, this furthermore implies that the helix can be formed only when spins take on a preferred value. This is exactly the situation we alluded to above in the sense that there exists a preferred orientation of spin. The situation with stretch-induced transformation from B-DNA to a more extended S-DNA is in fact similar since the formalism of the description is based on models of the helix-coil type \cite{Bruinsma, Punkkinen}. 

A preferred direction of the spin is not the only feature differentiating between different models. There is also the range of interactions that one needs to consider. Specifically, while in the case of a polypeptide chain it is essential that three successive spins all be in a chosen conformation corresponding to a single helix-inducing hydrogen bond \cite{cantor}, the description of DNA over-stretching implies that up to ten successive spins be engaged in a double helix-engendering hydrogen bond between opposing strands \cite{physa}. It thus transpires from both examples that it is necessary to consider some finite range of interaction and thus a finite number of nearest spins, $\Delta$, as crucial for the local formation of an ordered state -- a hydrogen bond in the considered case. 

There are thus two distinguishing features of the one-dimensional spin models worthy of further consideration. One is the number of nearest spins, entering into the local formation of an ordered state, and the other one is the existence of a preferred spin orientation pertinent to the ordered state. The effects of the former can be analyzed within the many-valued (Potts) spin model with an arbitrary but finite range of many-body interactions, while the latter forms the basis of {\sl The Generalized Model of Polypeptide Chain} (GMPC), accounting for the preferred spin orientation. GMPC has been formulated several decades ago \cite{anan,biopoly1,biopoly2} and has been extensively studied specifically in the context of the helix-coil transition \cite{bad10,bad11,bad12}. It was shown that the Zimm-Bragg model \cite{zb}  and the Lifson-Roig model \cite{lr} both correspond to particular cases of the GMPC variety with $\Delta=2$ and $\Delta=3$, respectively \cite{biopoly2,bad10}. The Wako-Saito-Munoz-Eaton (WSME) model, widely applied to protein folding (see \cite{WS,ME,bruscolini}), can also be shown to be related to the GMPC model (we will further comment on this point in the Conclusions). There is thus a spectrum of models that fall within the same class as the GMPC model. On the other hand, if no preferred spin value is taken into consideration, the standard Potts model with nearest-neighbor interactions ($\Delta=2$) \cite{baxter,wu} can be applied to a helix-coil transition, as shown by Goldstein \cite{goldstein}. However, for $\Delta>2$ there do not seem to exist any known solutions of the one-dimensional Potts model. We thus embark on a detailed study and comparison of the solutions of the two models for different values of $\Delta$, in order to connect the Potts and the GMPC model with many-body interactions with the models that do not allow for a spin Hamiltonian description but have been traditionally used to describe the statistical characteristics of the helix-coil transition \cite{zb,lr,polsher}.

\section{Helix-coil model formulation in terms of spin variables}
\label{sec:hcspin}

\begin{figure}[tbp]
\includegraphics[scale=0.6]{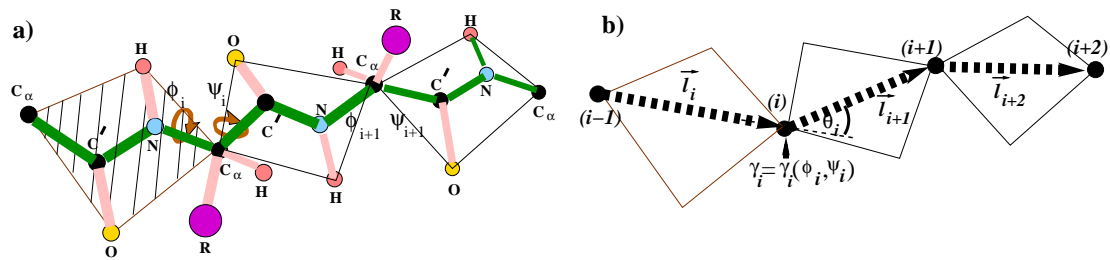}
\caption{
\label{ff1} (COLOR ONLINE) A polypeptide chain in a \emph{trans} conformation. Parallelograms indicate the plane of a virtual peptide bond. a) Schematic view of a polypeptide chain where the main-chain atoms are represented as rigid peptide segments, linked by virtual bonds through the $C_\alpha$ atoms. Each segment has two degrees of freedom due to the rotation around the $C_\alpha-C^\prime$ (torsional angle $\phi$) and $N-C_\alpha$ (torsional angle $\psi$) bonds. $R$ stands for the amino acid residues, while all other atoms have the corresponding chemical labels. b) Coarse-grained representation of a polypeptide chain: the conformation of ($i$)-th repeating unit is described with the help of bond length $l_i$, bond angle $\theta_i$ and a pair of torsional angles $\phi_i,\:\psi_i$. }
\end{figure}

Statistical description of polypeptide chain conformations involves important coarse-graining on the level of the $C_\alpha$ atoms, because of the planar configurations of the atomic groups (C$_{i-1}^{(\alpha )}$, C$_{i-1}$, O$_{i-1}$, N$_{i}$) due to specific bond hybridizations. The planar structure of these groups (peptide groups) allows the introduction of virtual bonds, connecting the neighboring asymmetric carbon atoms (Fig.~\ref{ff1}(a)) \cite{cantor,molbiol}. The configuration of a polypeptide chain can then be described with the sequence of (virtual) bond vectors $\{\vec{l}_i\}$ \cite{flory,dash,grosberg}, related to its backbone. In this description the bond lengths $\{l_i\}$, $i=1...N-1$, bond (valence) angles $\{\theta_i\}$, $i=1...N-2$, and pairs of $\{\phi_i, \psi_i\}$ torsional angles can be associated with each repeating unit. This description can be further simplified by taking into account that bond lengths and angle values usually vary within very narrow intervals (average fluctuation of $\pm 3 \div 5 \%$ at room temperature) and their fluctuations can be ignored \cite{flory}. The only relevant variables remaining are thus the $\{\phi_i, \psi_i\}$ torsional angles (Fig.~\ref{ff1}(b)). 

The conformational partition function of the repeating unit can be represented as a finite sum after discretization of the torsional angles (approximation of rotational isomers) \cite{volk}, opening up a possibility for a spin-based description of the polypeptide conformations. Assume that spin $\gamma_i$ describing the conformation of the $i$th repeating unit can take one of the $Q(\geq 2)$ values; $\gamma_i=1$ corresponding to values of the torsional angles $\{\phi_i, \psi_i\}$ from the helical region of the Ramachandran map, while the other $Q-1$ values correspond to torsional angles from allowed (not helical) region. The magnitude of $Q$ (number of spin orientations) can be identified with the ratio of the allowed region area versus helical region area on a Ramachandran map. According to the polypeptide chain geometry the equilibrium hydrogen bond formation can be established between the NH and CO groups, separated by three asymmetric carbon atoms \cite{cantor}; the energy $U$ is associated with every formed hydrogen bond corresponding to a {\sl coupling constant} $W=\exp{(U/T)}$, where $T$ is the temperature. One hydrogen bond thus restricts three \{$\phi ,\psi $\} pairs of rotation angles and establishes the structure with screw symmetry ($\alpha$-helix) \cite{poling}. Within the spin language this means that the hydrogen bond fixes three successive spin values along the chain. On the other hand, hydrogen bonds in double stranded DNA are formed between repeating units on the opposite strands and are approximately perpendicular to the DNA axis. Creation of hydrogen bonds in one pair of opposing bases thus applies restrictions to conformational states of $\sim 10$ neighbors (on the scale of single-strand Kuhn length) \cite{physa}. It makes sense to generalize and consider that one hydrogen bond formation restricts arbitrary (but finite) $\Delta$ number of spins \cite{biopoly2}, corresponding to many-body interactions. As helix formation comes at an entropic cost \cite{grosberg}, the larger is $\Delta$, the higher is such an entropic cost \cite{biopoly2}. The transformation from a coil to a helical conformation is energetically favorable (negative hydrogen bond energy is gained) but entropically unfavorable (the number of micro-states, available for repeating unit in a helical macro-state is decreased, as compared to the coil). As we show below, the compensation of energetic and entropic costs engenders a transition at the temperature corresponding to $\exp(U/T)=Q$.

\begin{figure}[tbp]
\includegraphics[scale=0.6]{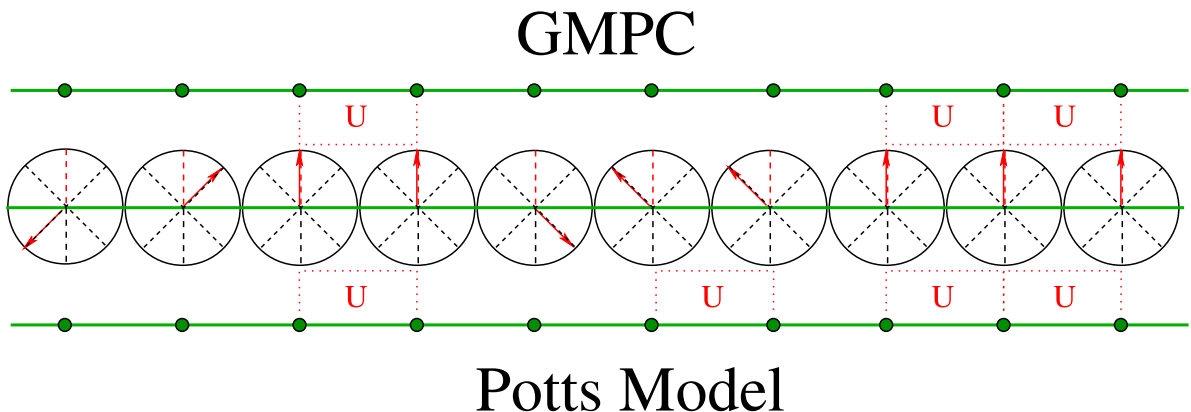}
\caption{
\label{ff2} (COLOR ONLINE) Schematic representation of a 10-mer with spins assigned to each repeating unit. Spins are shown as red arrows with $Q=8$ possible orientations. (Above) Assigned energy ($U$) in the nearest-neighbor ($\Delta=2$) GMPC model (with red dotted line indicating the preferred orientation of the spin). (Below) Potts model as in Ref. \cite{goldstein}. The Potts energy assignment results in a higher energy of the sample spin sequence.}
\end{figure}

To summarize, statistical description of the helix-coil transition requires three basic parameters: an energetic parameter, $W=\exp(U/T)$, where $U$ is the energy of a hydrogen bond; an entropic parameter, $Q$, that stands for the number of spin values; and a geometric parameter, $\Delta$, that describes the many-body geometry of the hydrogen bond formation. 
The corresponding Hamiltonian can thus be built in terms of the $\gamma_i$ spins \cite{anan,biopoly1,biopoly2, bad10,bad11,bad12} and corresponds to the GMPC model if the proper helix formation demands that $\Delta$ successive $\gamma$'s are all in the same preferred conformation number, {\sl e.g.} $1$ (see Fig.~\ref{ff2}, top). In the case of no preferred spin assignment to the helix formation we are then back to the Potts model (see Fig.~\ref{ff2}, bottom). In what follows we consider both types of spin Hamiltonians and discuss similarities and differences between the ensuing thermodynamics.

\section{Generalized model of polypeptide chain in transfer-matrix formalism}
\label{sec:gmpc}

\subsection{Hamiltonian}

The Hamiltonian for the Generalized Model of Polypeptide Chain (GMPC) is defined as

\begin{equation}
-\beta H=J\sum\limits_{i=1}^{N}\delta (\gamma _{i-2},1)\delta (\gamma
_{i-1},1)\delta (\gamma _{i},1),  \label{ham0}
\end{equation}

\noindent where $J=U/T$ is the reduced energy of the hydrogen bond. The strength of the hydrogen bond is between the valence bond and the van der Waals interactions. By definition the energy of hydrogen bond formation is negative. $\delta (a,b)$ is the Kronecker symbol. When generalized to any finite $\Delta$, the above Hamiltonian assumes the form

\begin{equation}
-\beta H=J\sum\limits_{i=1}^{N}\prod_{k=\Delta -1}^{0}\delta (\gamma
_{i-k},1)=J\sum\limits_{i=1}^{N}\delta _{i}^{(\Delta )},  \label{hamdelta}
\end{equation}
\noindent where $\delta _{i}^{(\Delta )}$ is the product of $\Delta $ Kronecker symbols for neighboring repeating units. 

\subsection{Transfer Matrix Approach and characteristic equation}

A transfer-matrix can be constructed corresponding to the Hamiltonian Eq.~\ref{hamdelta} for $\Delta =2$, $\Delta =3$ and $\Delta =4$ cases and in fact any finite $1<Q<\infty$. The algorithm for larger values of $\Delta$ is similar to that for the $\Delta =3$ and $\Delta =4$ cases.

Starting at $\Delta =2$ it is straightforward to show that the $Q \times Q$ transfer-matrix reads

\begin{equation}
\widehat{G(2)}=\left(
\begin{array}{cccccccccccccccc}
e^J & 1 & ... & 1\\
1 & 1 & ... & 1\\
... & ... & ... & ...\\
1 & 1 & ... & 1
\end{array}
\right),  \label{G2}
\end{equation}
\noindent and contains many ($Q-1$) identical rows and columns. At $\Delta =3$ the situation is more complicated, since many-body interactions began to play a role, and straightforward construction of the transfer-matrix is impossible. However, there is an efficient trick for the transfer-matrix construction. Instead of the three spin variables $\gamma _{i-2}, \gamma _{i-1}, \gamma _{i},$ one introduces a pair of two-index variables $\Omega _{i-1}=(\gamma _{i-1}^{\prime}, \gamma _{i})$ and $\Omega _{i}=(\gamma _{i-1},\gamma _{i})$ and sets to zero all elements of the matrix $\left( Q^{2}\times Q^{2}\right) $ for which $\gamma_{i-1}^{^{\prime }}\neq \gamma _{i-1}$ \cite{flory}. In other words, we write the Hamiltonian in the form

\begin{eqnarray}
-\beta H &=&J\sum\limits_{i=1}^{N}\delta (\gamma _{i-2},1)\delta (\gamma
_{i-1},1)\delta (\gamma _{i-1}^{^{\prime }},1)\delta (\gamma
_{i},1)= \label{H3} \\ &=&J\sum\limits_{i=1}^{N}\delta (\Omega
_{i-1},1)\delta (\Omega _{i},1),
\end{eqnarray}

\bigskip with the statistical weight 

\begin{equation}
g(\Omega _{i-1}\Omega _{i})=e^{-\beta H}\delta (\gamma
_{i-1}^{^{\prime }}\gamma _{i-1}).  \label{ves}
\end{equation}
\noindent Then the ($Q^2 \times Q^2$) transfer matrix reads:

\begin{equation}
\fl \widehat{G(3)}=\left(
\begin{array}{cccccccccccccccc}
e^J & 1 & ... & 1 & 0 & 0 & ... & 0 & ... & ... & ... & ... & 0 & 0 & ...
& 0 \\
0 & 0 & ... & 0 & 1 & 1 & ... & 1 & ... & ... & ... & ... & 0 & 0 & ... & 0
\\
... & ... & ... & ... & ... & ... & ... & ... & ... & ... & ... & ... & ...
& ... & ... & ... \\
0 & 0 & ... & 0 & 0 & 0 & ... & 0 & ... & ... & ... & ... & 1 & 1 & ... & 1
\\
1 & 1 & ... & 1 & 0 & 0 & ... & 0 & ... & ... & ... & ... & 0 & 0 & ... & 0
\\
0 & 0 & ... & 0 & 1 & 1 & ... & 1 & ... & ... & ... & ... & 0 & 0 & ... & 0
\\
... & ... & ... & ... & ... & ... & ... & ... & ... & ... & ... & ... & ...
& ... & ... & ... \\
0 & 0 & ... & 0 & 0 & 0 & ... & 0 & ... & ... & ... & ... & 1 & 1 & ... & 1
\\
... & ... & ... & ... & ... & ... & ... & ... & ... & ... & ... & ... & ...
& ... & ... & ... \\
... & ... & ... & ... & ... & ... & ... & ... & ... & ... & ... & ... & ...
& ... & ... & ... \\
... & ... & ... & ... & ... & ... & ... & ... & ... & ... & ... & ... & ...
& ... & ... & ... \\
... & ... & ... & ... & ... & ... & ... & ... & ... & ... & ... & ... & ...
& ... & ... & ... \\
1 & 1 & ... & 1 & 0 & 0 & ... & 0 & ... & ... & ... & ... & 0 & 0 & ... & 0
\\
0 & 0 & ... & 0 & 1 & 1 & ... & 1 & ... & ... & ... & ... & 0 & 0 & ... & 0
\\
... & ... & ... & ... & ... & ... & ... & ... & ... & ... & ... & ... & ...
& ... & ... & ... \\
0 & 0 & ... & 0 & 0 & 0 & ... & 0 & ... & ... & ... & ... & 1 & 1 & ... & 1
\end{array}
\right).  \label{G_3}
\end{equation}
\noindent Here again, there are many ($Q^2-\Delta$) identical rows and columns.  At $\Delta =4$ there are 4 spin variables and we add another 2, following a similar trick as above, so that a pair of three-index variables reads $\Omega _{i-1}=(\gamma _{i-3},\gamma _{i-2}^{\prime},\gamma _{i-1}^{\prime})$ and $\Omega _{i}=(\gamma _{i-2},\gamma _{i-1},\gamma _{i})$. The statistical weight is then prescribed as
\begin{equation}
g(\Omega _{i-1}\Omega _{i})=e^{-\beta H}\delta (\gamma_{i-2}, \gamma _{i-2}^{\prime})\delta (\gamma_{i-1}, \gamma _{i-1}^{\prime}).  
\label{ves4}
\end{equation}
The resulting transfer matrix has dimensions ($Q^3 \times Q^3$) and is by its structure similar to Eq.~\ref{G_3}. For larger $\Delta$s it is necessary to group $\gamma$s into $\Omega _{i-1}$ and $\Omega _{i}$ in a similar way. The procedure can be generalized and the proper statistical weight would then be written as

\begin{equation}
g(\Omega _{i-1},\Omega_{i})=e^{-H/T}\overset{1}{\underset{k=\Delta -2}{\prod }}\delta\left( \gamma _{i-k}^{^{\prime }},\gamma _{i-k}\right),
\label{vesdelta}
\end{equation}
\noindent resulting in a transfer matrix $\widehat{G(\Delta)}$ of dimensions ($Q^{\Delta-1} \times Q^{\Delta-1}$). Since there are ($Q^{\Delta-1}-\Delta$) identical rows and columns, the characteristic equation for $\widehat{G(\Delta)}$ turns out to be quite simple
\begin{equation}
\fl P_{GMPC}(\lambda,W,Q,\Delta)=\lambda^{Q^{\Delta-1}-\Delta}\{\lambda^\Delta-(W-1+Q)\lambda^{\Delta-1}+(W-1)(Q-1)\sum_{k=2}^{\Delta}\lambda^{\Delta-k}\}=0.
\label{chareq}
\end{equation}
\noindent It is obvious, that there are $\Delta$ non-trivial eigenvalues, so that to construct the thermodynamics of the model, it is enough to consider a transfer matrix of a much smaller, $(\Delta \times \Delta)$ size. Such a matrix has been derived in \cite{anan} by doing elementary transformation over $\widehat{G_{\Delta}}$ and looks like
\begin{equation}
\widehat{g}(\Delta)=\left(
\begin{array}{cccccc}
W-1 & W-1 & ... & W-1 & W-1 & W-1 \\
1 & 0 & ... & 0 & 0 & 0 \\
0 & 1 & ... & 0 & 0 & 0 \\
... & ... & ... & ... & ... & ... \\
0 & 0 & ... & 1 & 0 & 0 \\
0 & 0 & ... & 0 & 1 & Q
\end{array}
\right) . 
\label{gdelt}
\end{equation}
One can construct this transfer-matrix in the following way:
\begin{itemize}
\item{All elements of first row are equal to $W-1=e^J-1$;}
\item{All elements of the first lower pseudo-diagonal are $1$;}
\item{The element $\left( \Delta,\Delta \right) $ is $Q$;}
\item{All other elements are zero.}
\end{itemize}
\noindent Alternatively, elementary transformations can lead to 

\begin{equation}
\widehat{g^*}(\Delta)=\left(
\begin{array}{ccccccc}
W & 1 & 0 &... & 0 & 0 & 0 \\
0 & 0 & 1 &... & 0 & 0 & 0 \\
... & ... & ... & ... & ... & ... & ... \\
0 & 0 & 0 & ... & 0 & 1 & 0 \\
0 & 0 & 0 & ... & 0 & 0 & Q-1 \\ 
1 & 1 & 1 & ... & 1 & 1 & Q-1
\end{array}
\right) . 
\label{gdeltast}
\end{equation}
\noindent Both $\widehat{g}(\Delta)$ and $\widehat{g^*}(\Delta)$ have much smaller size than $\widehat{G(\Delta)}$ and result in the same characteristic equation
\begin{equation}
\fl p_{GMPC}(\lambda,W,Q,\Delta)=\lambda^\Delta-(W-1+Q)\lambda^{\Delta-1}+(W-1)(Q-1)\sum_{k=2}^{\Delta}\lambda^{\Delta-k}=0.
\label{chareqnt}
\end{equation}
\noindent By adding artificial $\lambda=1$ root, the characteristic equation can be written in much more compact form
\begin{equation}
\fl p_{GMPC}(\lambda,W,Q,\Delta)=\lambda^{\Delta-1}(\lambda-W)(\lambda-Q)-(W-1)(Q-1)=0.
\label{chareqbasic}
\end{equation}

\section{One dimensional Potts model with many-body interactions in transfer-matrix formalism}
\label{sec:potts}

\subsection{Hamiltonian}
Following Goldstein's formulation \cite{goldstein}, {\sl viz.} without any distinction between spin values, we construct the Hamiltonian for  a Potts model with $\Delta$ many-body interactions as
\begin{equation}
-\beta H=J\sum\limits_{i=1}^{N}\prod_{k=\Delta -1}^{1}\delta (\gamma_{i-k},\gamma_{i-k-1}).  
\label{hampottsdelta}
\end{equation}
\noindent One can notice, that for the same $\Delta$ many-body interactions the Hamiltonian of Potts model contains the product of $\Delta-1$ Kronecker symbols instead of exactly $\Delta$ such symbols as in the case of the GMPC model. This fact has important consequences, as we will show below. 

\subsection{Transfer Matrix Approach and characteristic equation}

The transfer-matrix corresponding to the Hamiltonian Eq.~\ref{hampottsdelta} can be constructed {\sl seriatim} for $\Delta =2$, $\Delta =3$, $\Delta =4$ and then for any finite $2<Q<\infty$. The algorithm for larger values of $\Delta$ is similar to that for the $\Delta =3$ and $\Delta =4$ cases.

At $\Delta =2$ it is straightforward to show that the $Q \times Q$ transfer-matrix reads

\begin{equation}
\widehat{G(2)}_{Potts}=\left(
\begin{array}{cccccccccccccccc}
e^J & 1 & ... & 1 \\
1 & e^J & ... & 1 \\
... & ... & ... & ...\\
1 & 1 & ... & e^J
\end{array}
\right).  \label{G2potts}
\end{equation}

At $\Delta =3$ we use the same trick used above for construction of transfer matrix of the GMPC model. Instead of the three spin variables $\gamma _{i-2}, \gamma _{i-1}, \gamma _{i},$ we now introduce a pair of two-index variables $\Omega _{i-1}=(\gamma _{i-2},\gamma _{i-1}^{\prime})$ and $\Omega _{i}=(\gamma _{i-1},\gamma _{i})$ and set to zero all elements of the matrix $\left( Q^{2}\times Q^{2}\right) $ for which $\gamma_{i-1}^{^{\prime }}\neq \gamma _{i-1}$ \cite{flory}. This results in transfer matrix
\begin{equation}
\fl \widehat{G(3)}_{Potts}=\left(
\begin{array}{cccccccccccccccc}
e^J & 1 & ... & 1 & 0 & 0 & ... & 0 & ... & ... & ... & ... & 0 & 0 & ...& 0 \\
0 & 0 & ... & 0 & 1 & 1 & ... & 1 & ... & ... & ... & ... & 0 & 0 & ... & 0 \\
... & ... & ... & ... & ... & ... & ... & ... & ... & ... & ... & ... & ...& ... & ... & ... \\
0 & 0 & ... & 0 & 0 & 0 & ... & 0 & ... & ... & ... & ... & 1 & 1 & ... & 1 \\
1 & 1 & ... & 1 & 0 & 0 & ... & 0 & ... & ... & ... & ... & 0 & 0 & ... & 0 \\
0 & 0 & ... & 0 & 1 & e^J & ... & 1 & ... & ... & ... & ... & 0 & 0 & ... & 0 \\
... & ... & ... & ... & ... & ... & ... & ... & ... & ... & ... & ... & ...& ... & ... & ... \\
0 & 0 & ... & 0 & 0 & 0 & ... & 0 & ... & ... & ... & ... & 1 & 1 & ... & 1 \\
... & ... & ... & ... & ... & ... & ... & ... & ... & ... & ... & ... & ...& ... & ... & ... \\
... & ... & ... & ... & ... & ... & ... & ... & ... & ... & ... & ... & ...& ... & ... & ... \\
... & ... & ... & ... & ... & ... & ... & ... & ... & ... & ... & ... & ...& ... & ... & ... \\
... & ... & ... & ... & ... & ... & ... & ... & ... & ... & ... & ... & ...& ... & ... & ... \\
1 & 1 & ... & 1 & 0 & 0 & ... & 0 & ... & ... & ... & ... & 0 & 0 & ... & 0 \\
0 & 0 & ... & 0 & 1 & 1 & ... & 1 & ... & ... & ... & ... & 0 & 0 & ... & 0 \\
... & ... & ... & ... & ... & ... & ... & ... & ... & ... & ... & ... & ...& ... & ... & ... \\
0 & 0 & ... & 0 & 0 & 0 & ... & 0 & ... & ... & ... & ... & 1 & 1 & ... & e^J
\end{array}
\right).  
\label{G_3potts}
\end{equation}
\noindent At $\Delta =4$ there are 4 spin variables and we add another 2, as above. The statistical weight is prescribed according to
\begin{equation}
g(\Omega _{i-1}\Omega _{i})=e^{-\beta H_{Potts}}\delta (\gamma_{i-2}, \gamma _{i-2}^{\prime})\delta (\gamma_{i-1}, \gamma _{i-1}^{\prime}).  
\label{ves4potts}
\end{equation}
The resulting transfer matrix has dimensions ($Q^3 \times Q^3$) and is by its structure similar to Eq.~\ref{G_3potts}. For larger $\Delta$s it is necessary to group $\gamma$s into $\Omega _{i-1}$ and $\Omega _{i}$ accordingly. The procedure can be generalized and the statistical weight would be written as 
\begin{equation}   
g(\Omega _{i-1},\Omega_{i})=e^{-H/T}\overset{1}{\underset{k=\Delta -2}{\prod }}\delta\left( \gamma _{i-k}^{^{\prime }},\gamma _{i-k}\right),
\label{vesdeltapotts}
\end{equation}
\noindent resulting in a transfer matrix $\widehat{G(\Delta)}_{Potts}$ of dimensions ($Q^{\Delta-1} \times Q^{\Delta-1}$). This matrix differs from the GMPC case (see Eq.~\ref{G_3}) in that all the diagonal elements are multiplied by $e^J$, while in Eq.~\ref{G_3} only the element $(1,1)$ is multiplied by this factor. Here again, there are $a(Q,\Delta)=Q^{\Delta-1}-Q(\Delta-1)$ identical rows and columns. The corresponding characteristic equation then follows as 
\begin{eqnarray}
\label{chareqpotts}
\fl P_{Potts}(\lambda,W,Q,\Delta) = \\ \fl \lambda^{a(Q,\Delta)}\left(\lambda^{\Delta-1}-(W-1+Q)\lambda^{\Delta-2}+(W-1)(Q-1)\sum_{k=2}^{\Delta-1}\lambda^{\Delta-1-k}\right) \times \\ \fl \left(\lambda^{\Delta-1}-(W-1)\lambda^{\Delta-2}-(W-1)\sum_{k=2}^{\Delta-1}\lambda^{\Delta-1-k}\right)^{Q-1}=0.
\end{eqnarray}
\noindent Elimination of trivial eigenvalues results in
\begin{eqnarray}
\label{chareqpotts1}
\fl p_{Potts}(\lambda,W,Q,\Delta) = \\ \fl \left(\lambda^{\Delta-1}-(W-1+Q)\lambda^{\Delta-2}+(W-1)(Q-1)\sum_{k=2}^{\Delta-1}\lambda^{\Delta-1-k}\right)\times \\ \fl \left(\lambda^{\Delta-1}-(W-1)\lambda^{\Delta-2}-(W-1)\sum_{k=2}^{\Delta-1}\lambda^{\Delta-1-k}\right)^{Q-1}=0.
\end{eqnarray}
\noindent Unfortunately, it is not possible to derive a simpler transfer-matrix, that would correspond to such characteristic equation. We have checked Eq.~\ref{chareqpotts1} to be true up to $\Delta=7$ by hand and using Wolfram Mathematica software. 

\section{Comparison of characteristic equations for the GMPC and Potts models}
\label{sec:compare}

Since the transfer-matrix, being a matrix of statistical weights, is non-negative, Frobenius-Perron theorem applies, and there exists a positive, non-degenerate maximal eigenvalue $\lambda_1$. After solving the characteristic equation and assuming cyclic boundary conditions, we can straightforwardly reconstruct the partition function as
\begin{equation}
Z(\lambda)=\lim_{N \rightarrow \infty} \sum_{i=1}^\Delta \lambda_i^N = \lambda_1^N,
\label{partfunc}
\end{equation}
\noindent with the free energy as
\begin{equation}
F(\lambda)=-TN\ln \lambda_1,
\label{free}
\end{equation}
\noindent and the spatial correlation length as
\begin{equation}
\xi(\lambda)=\ln^{-1}\left(\frac{\lambda_1}{\lambda_2}\right),
\label{ksi}
\end{equation}
\noindent where $\lambda_1$ is the maximal and $\lambda_2$ is the second largest eigenvalue. 
This means that the thermodynamics of the model is determined by its characteristic equation. 

The comparison of Eq.~\ref{chareqnt} with Eq.~\ref{chareqpotts1} reveals similarities and differences between the two considered models, as obviously 
\begin{equation} \label{compare}
\fl p_{Potts}(\lambda,W,Q,\Delta)=p_{GMPC}(\lambda,W,Q,\Delta-1) \times p_{GMPC}(\lambda,W,Q=0,\Delta-1)^{Q-1},
\end{equation} 
\noindent and therefore the properties of the Potts model defined by $p_{Potts}(\lambda,W,Q,\Delta)$ are related to the properties of the GMPC model of $p_{GMPC}(\lambda,W,Q,\Delta-1)$. In the region of positive temperatures ($W>1$) the first bracket of Eq.~\ref{compare} has two positive roots, while the second bracket has a single, positive and $Q-1$ times degenerate root. 

\begin{figure*}[t!]\begin{center}
\begin{minipage}[b]{0.32\textwidth}\begin{center}
\includegraphics[width=\textwidth]{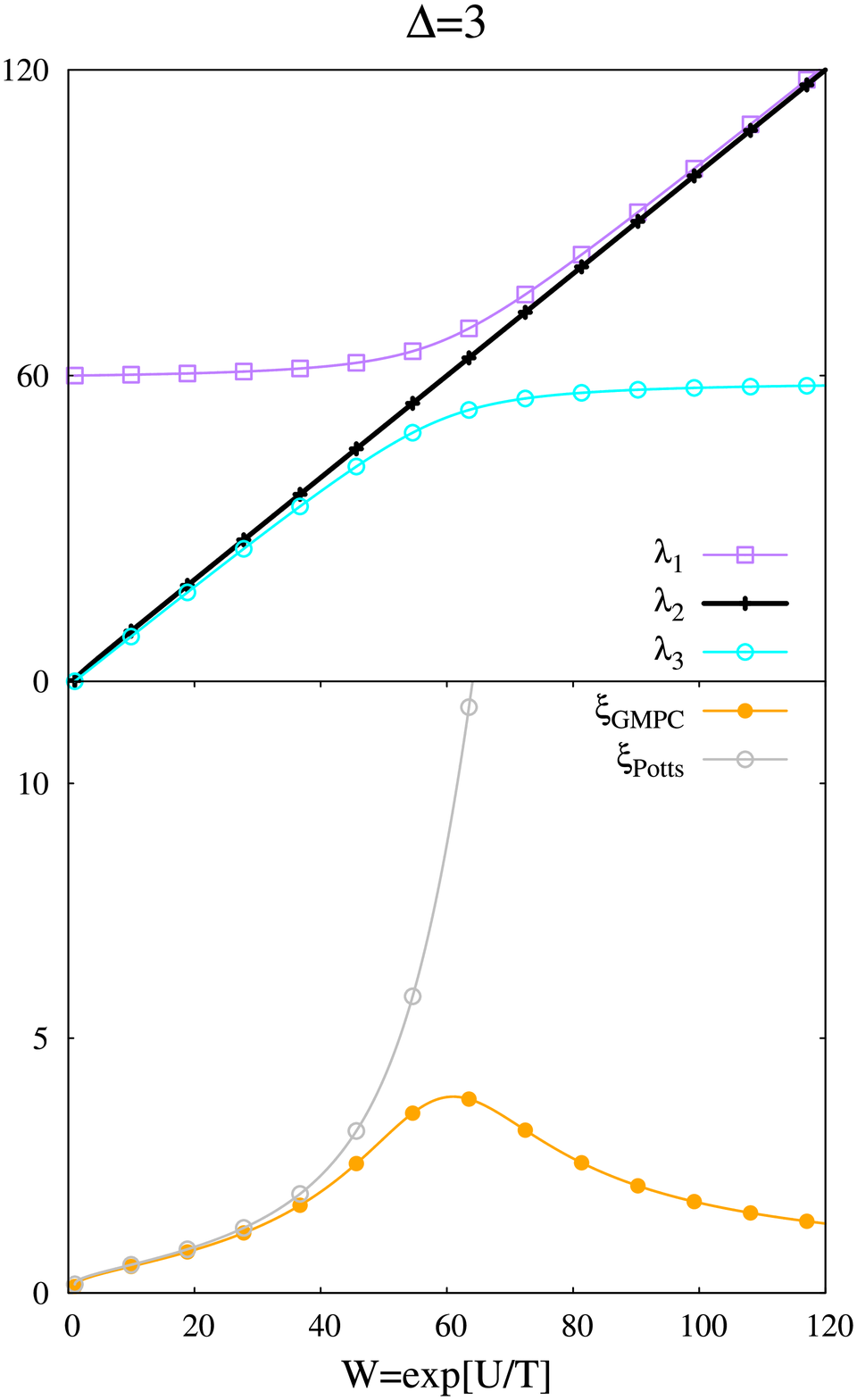} (a)
\end{center}\end{minipage}
\begin{minipage}[b]{0.32\textwidth}\begin{center}
\includegraphics[width=\textwidth]{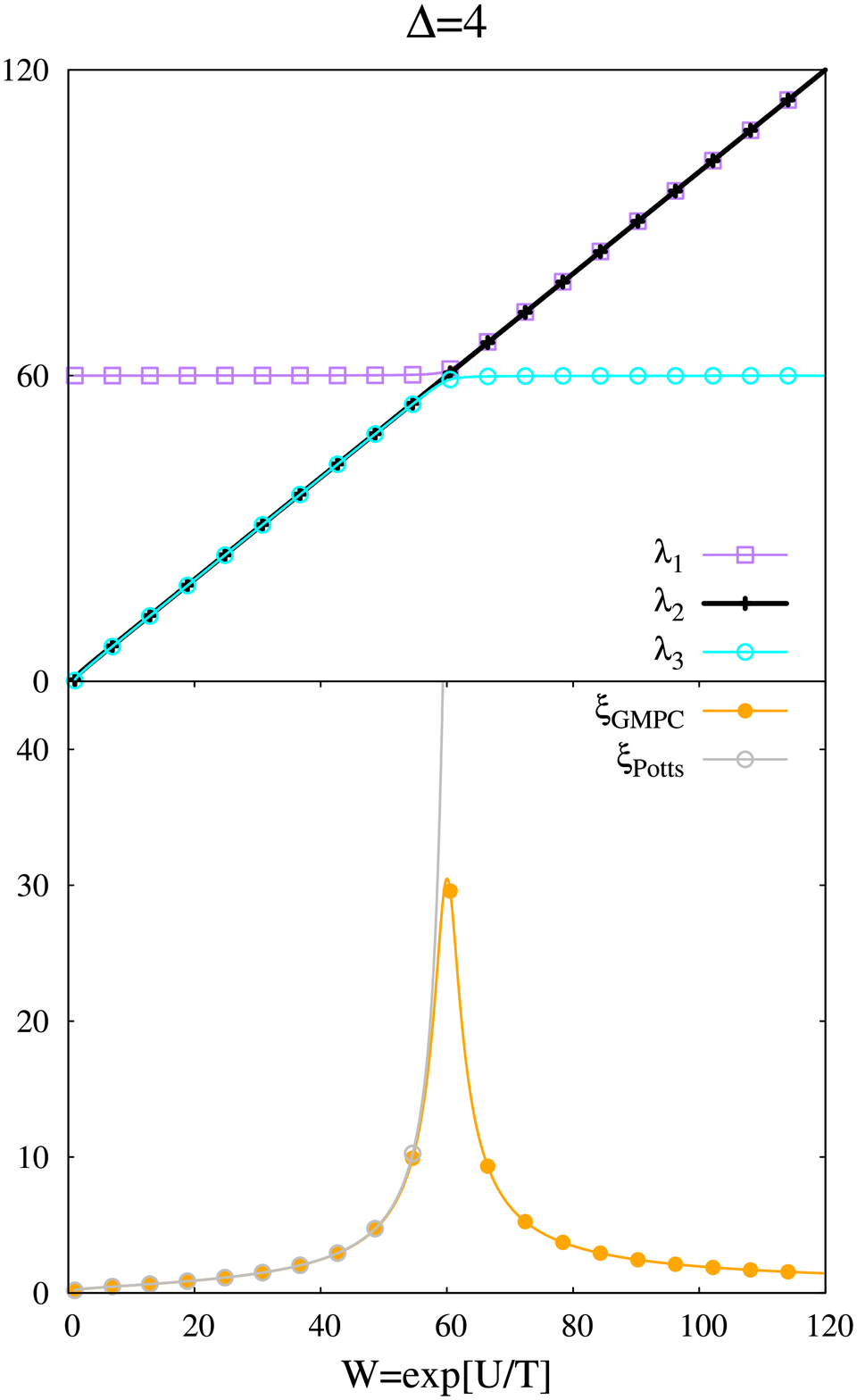} (b)
\end{center}\end{minipage}
\begin{minipage}[b]{0.32\textwidth}\begin{center}
\includegraphics[width=\textwidth]{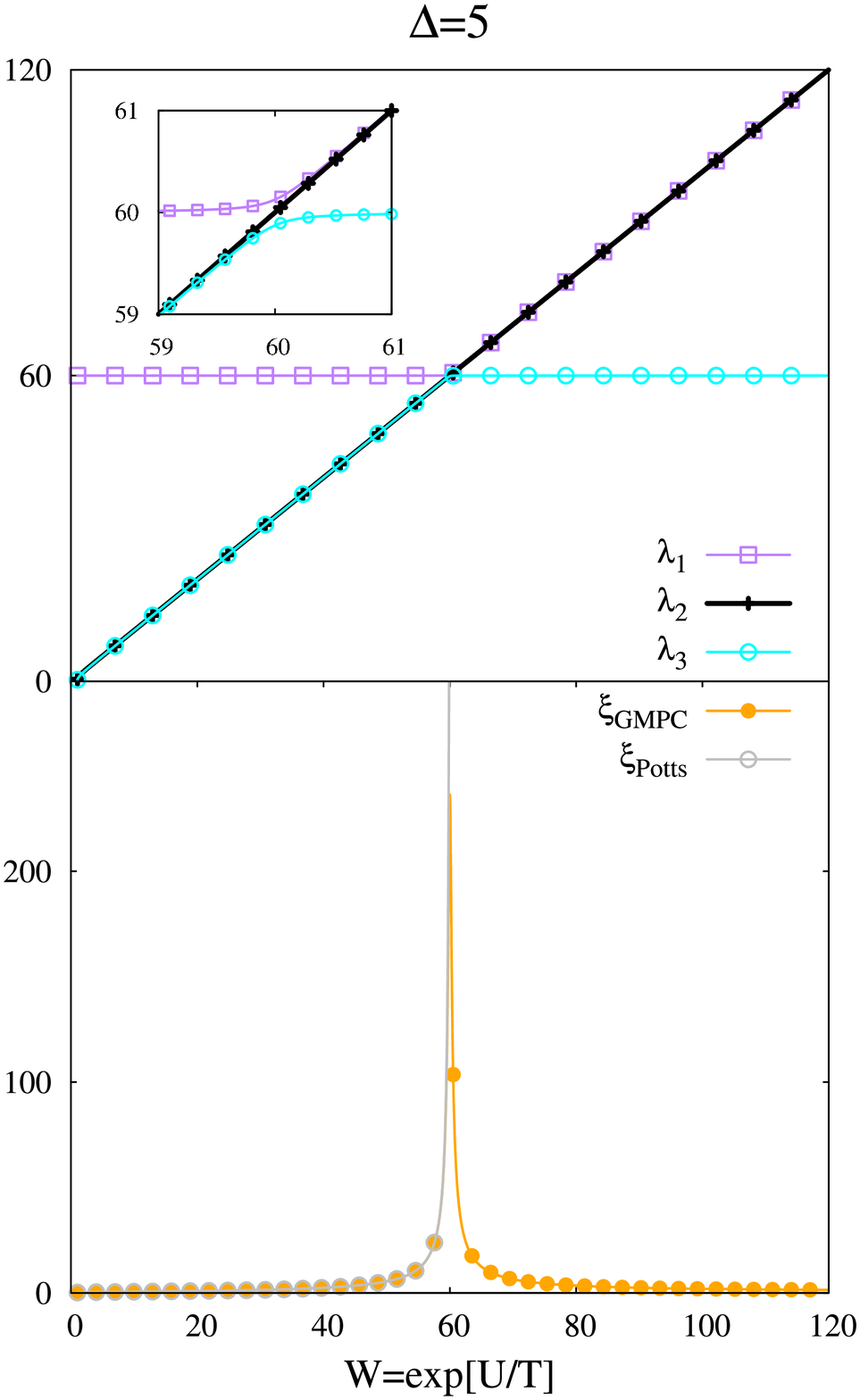} (c)
\end{center}\end{minipage}
\caption{COLOR ONLINE) (a) Three largest eigenvalues (top) from Eq.~\ref{compare} vs $W=\exp[U/T]$ for $\Delta=3$ and $Q=60$. Correlation lengths for $\Delta$ particle Potts and $\Delta-1$ particle GMPC models (bottom) (b) $\Delta=4$. (c) $\Delta=5$.}
\label{ff}
\end{center}\end{figure*}

Detailed analysis reveals that the root from the second bracket is always in between the two roots from the first bracket (see Fig.~\ref{ff}), so that the maximal root (which determines the free energy) always originates from the first bracket. In other words, the one-dimensional Potts model with $\Delta$ many-body interactions has the same free energy as $\Delta-1$-many-body GMPC model. All averages that depend on the largest eigenvalue only, such as helicity degree or number of junctions \cite{biopoly1,biopoly2}, are therefore the same for both models. 

However, correlation lengths are obviously different, as is illustrated on Fig.~\ref{ff}. This follows since the correlation length for the Potts model is determined {\sl via} $\lambda_1$ and $\lambda_2$, while for the GMPC model {\sl via} $\lambda_1$ and $\lambda_3$. Indeed, as one can see on Fig.~\ref{ff}, while the correlation lengths of both models coincide at low $W$ (high temperatures), they differ essentially at intermediate to high values of $W$ (low temperatures). The correlation length of the Potts model abruptly increases close to the temperature, where energetic and entropic parameters compensate each other ($W=Q$), while the correlation length of the GMPC model goes through a maximum at this point. The appearance of large correlations in the Potts model is a direct consequence of the absence of a preferred spin value, meaning that the standard Potts model is not suitable to describe the systems with a special, preferred direction of the spin, conditioned by an external field or by the structure of the system, as is the case in biopolymers. 

\section{Conclusion}
\label{sec:conclusion}

The one-dimensional nature of the considered models allowed us to explicitly construct the transfer-matrix for $\Delta$ many-body Potts model Eq.~\ref{G_3potts} and to derive the corresponding secular equation Eq.~\ref{chareqpotts}. The three largest eigenvalues, most important for resulting thermodynamics, were considered for several finite $\Delta$s. As a result, it was found out that only the free energy (largest eigenvalue) of $\Delta$ many-body Potts model is equal to that of the $\Delta-1$ many-body GMPC model, while the correlation lengths (which depend on the second largest eigenvalue as well) of two models differ significantly. 

The analysis presented above shows that a Potts model with $\Delta=2$ \cite{goldstein}, formulated on the level of the effective free energy, is equivalent to the one-body GMPC model ($\Delta=1$). Since the Zimm-Bragg model has been shown as originating from the $\Delta=2$ GMPC model \cite{bad10}, this means that in order to achieve at least the same level of description, the former approach should be extended to next-nearest neighbor, three-body, interactions. The characteristic equation of the Lifson-Roig model can be derived from GMPC model Hamiltonian with three-body interactions \cite{biopoly1}, so that the Potts model with $\Delta=4$ would be necessary to get an equivalent free energy. 

It is worth noticing that the GMCP model is also related to another interesting model that has been frequently used in the framework of protein folding, namely the 
Wako-Saito-Munoz-Eaton (WSME) model \cite{WS,ME,bruscolini}. Unlike the GMPC model, and quite similarly to the Zimm-Bragg model, the WSME model sets out from a phenomenological expression of the free energy. The methodology to pass from the GMPC Hamiltonian model to the corresponding free energy has already been elucidated in Ref.~\cite{bad10}, and it turns out that the resulting free energy bears strong similarities with the corresponding WSME one -- being in fact equivalent for finite range interactions, apart from an appropriate rescaling in the parameters. Correlation lengths are, however, in general different for essentially the same reasons given in 
the present work for the Potts model. As this point appears to be interesting, it will be the subject of a future dedicated analysis.


Though the derivation of the secular equation of the many-body Potts model is straightforward, it was not derived or analyzed before. Interesting differences in the thermodynamic behavior appear once one of the spin values is preferred, as in the GMPC model. In the case when the isotropy of spins is broken in this way, the application of the GMPC model as opposed to the Potts model seems to be more adequate, as shown explicitly for the example of the helix-coil transition. Since a fruitful analogy between magnetic and polymer systems is well established and long known, we believe that there are likewise situations in the theory of magnetism, where application of the GMPC model instead of the standard Potts model would lead to a more detailed understanding of the thermodynamic properties. 


{\ack
AB and AG acknowledge the support from PRIN-COFIN 2007 grant. RP and AB acknowledge ARRS grants P1-0055 and J1-4297.
}
\bibliographystyle{apsrev}

\end{document}